\documentclass[conference]{IEEEtran}

\usepackage{cite}

\usepackage[pdftex]{graphicx} % Use pdftex driver for pdfLaTeX

\usepackage{amsmath}
\usepackage{amssymb}
\usepackage{amsfonts}

\usepackage{array}

\usepackage{hyperref} 
\hypersetup{
    colorlinks=true,
    linkcolor=blue,
    filecolor=magenta,
    urlcolor=blue,
    citecolor=green,
    pdftitle={Enterprise-Grade Security for the Model Context Protocol (MCP): Frameworks and Mitigation Strategies},
    pdfpagemode=FullScreen,
    breaklinks=true % Added to help break long URLs
    }

\usepackage{booktabs}
\usepackage{tabularx} 

\begin{document}
\IEEEoverridecommandlockouts 
\title{Enterprise-Grade Security for the Model Context Protocol (MCP): Frameworks and Mitigation Strategies}

\author{\IEEEauthorblockN{Vineeth Sai Narajala \thanks{This work is not related to the author's position at Amazon Web Services.}}
\IEEEauthorblockA{Amazon Web Services\\
Email: vineesa@amazon.com}
\and
\IEEEauthorblockN{Idan Habler}
\IEEEauthorblockA{Adversarial AI Security reSearch (A2RS), Intuit\\
Email: idan\_habler@intuit.com}
}

\maketitle

\begin{abstract}
The Model Context Protocol (MCP), introduced by Anthropic, provides a standardized framework for artificial intelligence (AI) systems to interact with external data sources and tools in real-time. While MCP offers significant advantages for AI integration and capability extension, it introduces novel security challenges that demand rigorous analysis and mitigation. This paper builds upon foundational research into MCP architecture and preliminary security assessments to deliver enterprise-grade mitigation frameworks and detailed technical implementation strategies. Through systematic threat modeling and analysis of MCP implementations and analysis of potential attack vectors, including sophisticated threats like tool poisoning, we present actionable security patterns tailored for MCP implementers and adopters. The primary contribution of this research lies in translating theoretical security concerns into a practical, implementable framework with actionable controls, thereby providing essential guidance for the secure enterprise adoption and governance of integrated AI systems.
\end{abstract}

% keywords

\IEEEpeerreviewmaketitle

\section{Introduction}
% Introduction subsection
\subsection{Background: The Model Context Protocol (MCP)}

The Model Context Protocol (MCP) is a major step forward in standardizing how AI models interact with the world around them---giving them the ability to use tools and access real-time data on the fly \cite{anthropic2024mcp}. Foundational research, like the work by Hou et al. \cite{hou2025mcp}, has laid out the basic components, how MCP works, and some early security concerns. Building on that, our focus shifts to what it takes to secure MCP in real-world, enterprise-level deployments.

To understand the security implications of MCP, it is essential to understand its fundamental architecture, which comprises three key components:
\begin{itemize}
    \item \textbf{MCP Host:} Refers to the AI application or environment in which AI-driven tasks are performed and that operates the MCP client. Examples include applications such as Claude Desktop, and AI-driven integrated development tools like Cursor for software development. The host integrates tools and data, enabling interaction with external services through the MCP client and server.
    \item \textbf{MCP Client:} Serves as an intermediary in the host environment, facilitating communication between the MCP host and MCP servers. It sends requests and seeks information about the available services of servers. Data communication with servers is conducted securely and reliably via the transport layer.
    \item \textbf{MCP Server:} Serves as a gateway that allows the MCP client to interact with external services and execute tasks. The MCP Server offers three essential functionalities:
        \begin{itemize}
            \item \textit{Tools:} Allows the server to invoke external services and APIs to execute tasks on behalf of the AI model.
            \item \textit{Resources:} Exposes structured or unstructured datasets (e.g., local files, databases, and cloud platforms) to the model. When a model needs specific data (e.g., customer logs for a recommendation system), the server retrieves and processes it.
            \item \textit{Prompts:} Are reusable templates managed by the server to enhance model responses, maintain consistency, and simplify repeated actions.
        \end{itemize}
\end{itemize}

As MCP moves from theory into production, strong and scalable security becomes critical. Standard API security practices remain important but are insufficient to address the unique risks associated with MCP's dynamic, tool-based model. One example is "tool poisoning"—a type of attack where maliciously crafted tool descriptions trick AI models into doing things they shouldn't \cite{goodin2024poisoned}. Our goal in this paper is to go beyond identifying risks, offering actionable security strategies that enterprises can implement to safeguard their MCP systems.

\subsection{The Broader Context: MCP Security and AI Governance}

Securing protocols like MCP isn't just a technical issue---it's closely tied to the bigger picture of AI governance and responsible AI use. As AI systems become more autonomous and start interacting directly with external tools and real-time data through things like MCP, making sure those interactions are secure becomes absolutely essential. If an MCP implementation is compromised, the consequences can be severe—ranging from data leaks and unauthorized AI actions to loss of user trust and real-world harm, particularly when AI interacts with physical systems such as those in industrial IoT (IIoT). As a result, securing MCP is essential to building trustworthy AI systems, especially in complex enterprise environments. This paper contributes to that goal by providing practical strategies to mitigate the unique security risks introduced when AI systems interact with external tools.

\subsection{Research Focus and Contributions}

This paper moves beyond identifying potential MCP vulnerabilities to providing actionable security controls tailored for enterprise contexts. Our primary contributions are:
\begin{enumerate}
    \item A Comprehensive, Multi-Layered Security Framework: Detailing a defense-in-depth strategy specifically adapted for the nuances of MCP deployments.
    \item Actionable Zero Trust Implementation Patterns: Providing specific guidance on applying Zero Trust principles \cite{rose2020zero, kindervag2010zero} within MCP environments, focusing on identity, access, and continuous verification.
    \item Advanced Threat Mitigation Techniques: Outlining sophisticated validation and detection methods to counter tool poisoning, data exfiltration, and manipulation attacks.
    \item Operational Security Procedures: Defining best practices for ongoing monitoring, incident response, and threat intelligence integration specific to MCP.
    \item Secure Reference Architectures: Illustrating practical deployment patterns for secure MCP integration within enterprise ecosystems.
\end{enumerate}
While prior research \cite{hou2025mcp} has mapped the MCP landscape and identified potential threat categories, our work distinguishes itself by delivering detailed, actionable security controls and configurations designed for immediate application by security practitioners and system architects responsible for hardening MCP implementations.

\section{Security Threat Landscape and Methodology}

% --- BEGIN FIGURE 1 ---
\begin{figure*}[!t] 
\centering
\includegraphics[width=1\textwidth]{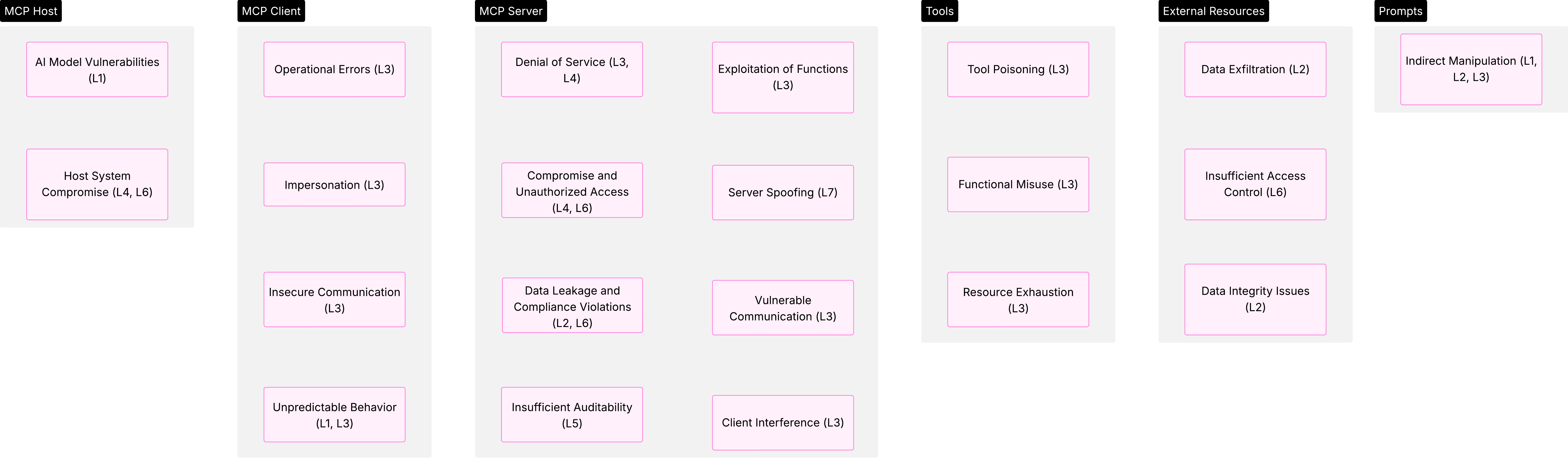}
\caption{This figure provides a comprehensive categorization of security threats across different components of the Model Context Protocol (MCP) architecture. The diagram organizes threats by component, with each component displayed as a column and specific threats listed within them.}
\label{fig:threat_categorization}
\end{figure*}
% --- END FIGURE 1 ---

\subsection{Threat Modeling Methodology for MCP}

Building upon our general risk identification approach \cite{securing_agentic_ai_2025}, we employ the MAESTRO framework \cite{csa2025maestro} for comprehensive threat modeling of AI systems as applied to MCP \cite{huang_threatguide}. This framework provides a systematic methodology by examining potential vulnerabilities across seven specific layers of an AI system's architecture, allowing for a structured analysis of the MCP security landscape.

\subsubsection{MAESTRO Framework for MCP}
The MAESTRO framework examines AI system vulnerabilities across seven distinct layers, each relevant to different aspects of MCP security:
\begin{itemize}
    \item L1 --Foundation Models: Concerns related to the underlying AI models, training data, and inherent vulnerabilities.
    \item L2 -- Data Operations: Security of external data management and integration within MCP systems.
    \item L3 --Agent Frameworks: Vulnerabilities in agent logic, protocols, and tool utilization mechanisms.
    \item L4 -- Deployment Infrastructure: Security of hardware and software environments hosting MCP components.
    \item L5 -- Evaluation \& Observability: Vulnerabilities in monitoring and diagnostic systems for MCP behavior.
    \item L6 -- Security \& Compliance: Issues related to access controls, policy enforcement, and regulatory requirements.
    \item L7 --Agent Ecosystem: Security challenges in interactions with humans, external tools, or other agents.
\end{itemize}
By applying this layered approach to MCP, we can systematically identify and categorize threats across the entire protocol stack, from underlying models to deployment infrastructure and ecosystem interactions. Figure~\ref{fig:threat_categorization} illustrates the comprehensive threat categories.

\subsubsection{Key Threat Categories by MCP Component}
Using the MAESTRO framework, we identified the following key threats organized by MCP component:

\textbf{MCP Server Threats}
\begin{itemize}
    \item Compromise and Unauthorized Access: Vulnerabilities in deployment or security configurations that permit attacker access (L4, L6).
    \item Exploitation of Functions: Vulnerable tools potentially utilized to execute harmful operations (L3).
    \item Denial of Service: Server flooding through excessive requests or autonomous agent loops (L3, L4).
    \item Vulnerable Communication: Interception or modification of data between client and server (L3).
    \item Client Interference: Insufficient isolation allowing one client's behavior to affect others (L3).
    \item Data Leakage and Compliance Violations: Unauthorized data exfiltration or regulatory violations (L2, L6).
    \item Insufficient Auditability: Inadequate logging restricting detection and investigation of security events (L5).
    \item Server Spoofing: Malicious servers posing as legitimate ones within the ecosystem (L7).
\end{itemize}

\textbf{MCP Client Threats}
\begin{itemize}
    \item Impersonation: Attackers posing as legitimate clients to gain unauthorized access (L3).
    \item Insecure Communication: Vulnerabilities leading to interception or data alteration during transmission (L3).
    \item Operational Errors: Schema inconsistencies between client and server causing functional errors (L3).
    \item Unpredictable Behavior: Model instability prompting disruptive or unexpected requests (L1, L3).
\end{itemize}

\textbf{MCP Host Environment Threats}
\begin{itemize}
    \item AI Model Vulnerabilities: Hallucinations, instability, or specific exploits resulting in erroneous MCP utilization (L1).
    \item Host System Compromise: General compromise of the host system affecting MCP client security (L4, L6).
\end{itemize}

\textbf{Data Sources and External Resources Threats}
\begin{itemize}
    \item Insufficient Access Control: Inadequate security measures on underlying data sources enabling unauthorized access (L6).
    \item Data Integrity Issues: Data drift or intentional tampering leading to misleading results (L2).
    \item Data Exfiltration: Unauthorized extraction of confidential information through compromised connections (L2).
\end{itemize}

\textbf{Tool-Related Threats}
\begin{itemize}
    \item Functional Misuse: Tools utilized for improper or harmful purposes beyond intended functionality (L3).
    \item Resource Exhaustion: Excessive tool usage overwhelming system resources (L3).
    \item Tool Poisoning: Malicious manipulation of tool descriptions or parameters to induce unintended model behavior (L3).
\end{itemize}

\textbf{Prompt-Related Threats}
\begin{itemize}
    \item Indirect Manipulation: Attacks targeting the foundation model (e.g., prompt injection), influencing MCP utilization (L1, L2, L3).
\end{itemize}
This threat analysis informs our comprehensive security framework and mitigation strategies presented in Section \ref{sec:framework}, enabling a defense-in-depth approach tailored to the specific security challenges of MCP implementations in enterprise environments.

\subsection{Key Security Challenges}

MCP introduces a distinct set of security challenges because it acts as a bridge between powerful AI models and often untrusted external tools and data sources. Its dynamic nature creates a complicated trust landscape that includes the MCP server, the AI model itself, client applications, and the various tools plugged into the system.

Building upon the threat categorization identified by Hou et al. \cite{hou2025mcp}---covering installation, operational, and update-related risks---our focus is on mitigating enterprise-critical threats, including:
\begin{itemize}
    \item Tool Poisoning: Malicious manipulation of tool descriptions or parameters to induce unintended or harmful actions by the AI model.
    \item Data Exfiltration: Unauthorized extraction of sensitive data through compromised tools or manipulated MCP responses.
    \item Command and Control (C2): Establishment of covert C2 channels via compromised MCP servers or tools.
    \item Update Mechanism Compromise: Insertion of persistent backdoors or malware through compromised MCP server or tool update channels.
    \item Identity and Access Control Subversion: Exploitation of authentication or authorization flaws to gain unauthorized access or escalate privileges.
    \item Denial of Service (DoS): Overloading MCP servers or dependent resources through excessive requests or resource exhaustion attacks.
    \item Insecure Configuration: Exploitation of misconfigurations in MCP servers, network settings, or access controls.
\end{itemize}

\subsection{Methodology}

The security framework and mitigation strategies presented in this paper were developed using the following methodology:
\begin{enumerate}
    \item Risk Identification and Analysis: We identified potential security risks through a combination of:
        \begin{itemize}
            \item Reviewing existing literature on MCP and related AI integration security challenges \cite{goodin2024poisoned}.
            \item Analyzing the MCP protocol specification and its interaction model to identify potential weak points.
            \item Applying established threat modeling concepts (drawing inspiration from frameworks like STRIDE - Spoofing, Tampering, Repudiation, Information Disclosure, Denial of Service, Elevation of Privilege) to the MCP ecosystem.
            \item Extrapolating threats from known vulnerabilities in analogous technologies like APIs, web services, and agent-based systems \cite{owasp2024agentic}.
            \item Conducting structured threat modeling of MCP architecture.
        \end{itemize}
        Prioritization of risks was based on qualitative assessment of potential impact and likelihood, informed by the attack scenarios and known vulnerability patterns in similar systems.
    \item Mitigation Strategy Development: We designed mitigation strategies by:
        \begin{itemize}
            \item Synthesizing established security best practices from recognized sources, including National Institute of Standards and Technology (NIST) Special Publications (SP) 800-53 Rev 5 \cite{nist2023adversarial} (Security Controls) and SP 800-207 \cite{rose2020zero} (Zero Trust Architecture), Open Web Application Security Project (OWASP) guidelines \cite{owasp2024agentic}, and general secure coding and threat modeling principles.
            \item Adapting these general principles specifically to the unique characteristics of MCP, such as the dynamic nature of tool interaction and the potential for semantic manipulation (tool poisoning).
            \item Structuring the mitigations around core security principles: defense-in-depth and Zero Trust.
            \item Organizing controls across multiple layers: network, application gateway, container/host, identity/access, tool/prompt management, input/output validation, and operations.
        \end{itemize}
\end{enumerate}

\section{Comprehensive MCP Security Framework} \label{sec:framework}

\begin{figure*}[!t] % Note the asterisk * for double column
\centering
\includegraphics[width=1.02\textwidth]{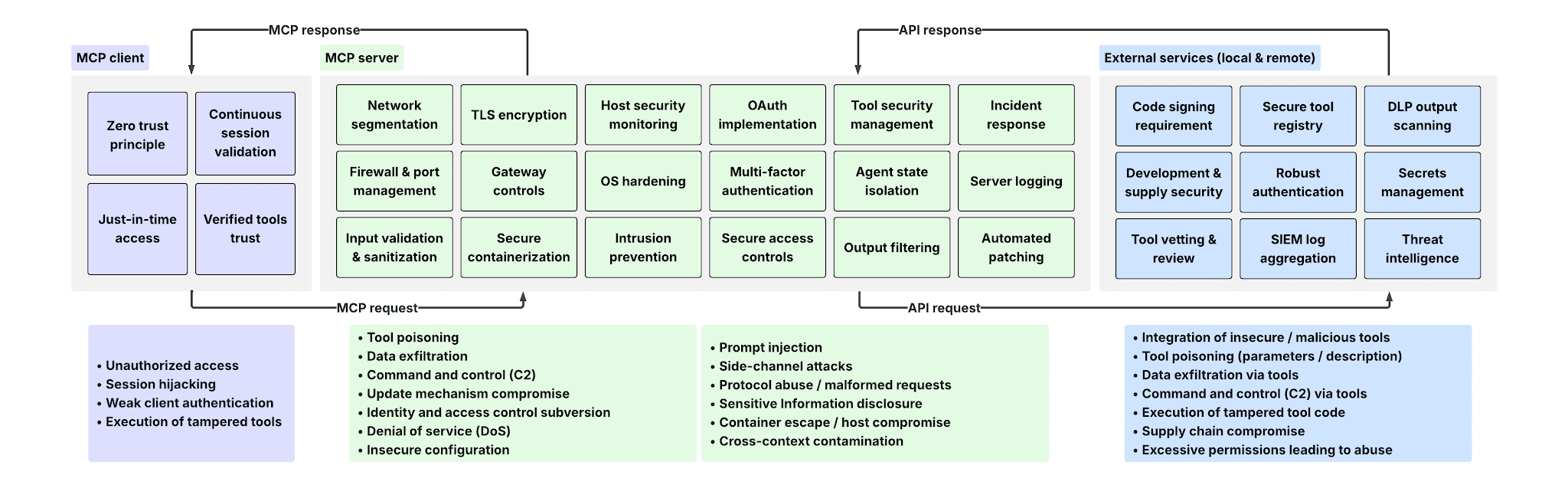} % Replace figure3.png with your framework image file, adjust width
\caption{This figure illustrates a multi-layered security framework for Model Context Protocol (MCP) based on defense-in-depth and Zero Trust principles.}
\label{fig:framework} % Kept label fig:framework
\end{figure*}

We propose a multi-layered security framework based on defense-in-depth and Zero Trust principles, tailored to the specific risks of MCP. (See Table \ref{tab:threats_mitigations} for a summary of key threats and corresponding mitigation controls). The overall security framework (see Figure~\ref{fig:framework}) provides a high-level overview of the security framework.

\subsection{MCP Server-Side Mitigations}

\subsubsection{Network Segmentation and Microsegmentation}
Network segmentation is a fundamental security strategy that goes beyond traditional perimeter-based defenses. In MCP environments, this approach is exponentially more critical due to the protocol's dynamic nature of tool interactions.
\begin{itemize}
    \item \textbf{Dedicated MCP Security Zones:} Isolate MCP servers and critical components within dedicated network segments (e.g., Virtual Local Area Networks (VLANs), Virtual Private Cloud (VPC) subnets) with strict ingress/egress filtering rules based on the principle of least privilege. Use Security groups as well in Cloud Environments like AWS.
    \item \textbf{Service Mesh Implementation:} Employ a service mesh (e.g., Istio) to enforce fine-grained, identity-based traffic control (mutual Transport Layer Security - mTLS) between MCP microservices and connected tools, independent of network topology when using Kubernetes architecture.
    \item \textbf{Application-Layer Filtering Gateways:} Deploy gateways (e.g., Web Application Firewalls (WAFs), API Gateways) capable of deep packet inspection (DPI) for MCP traffic, configured with rules to detect protocol anomalies, malicious payloads in tool descriptions/parameters, and known attack signatures.
    \item \textbf{End-to-End Encryption:} Mandate TLS 1.2+ with strong cipher suites and certificate pinning for all MCP-related communication paths.
\end{itemize}
\textit{Rationale:} Contains breaches, prevents lateral movement, and protects data in transit.

\subsubsection{Application Gateway Security Controls}
Application-level gateways inspecting MCP traffic should enforce:
\begin{itemize}
    \item \textbf{Strict Protocol Validation:} Rigorously validate all MCP messages against the official protocol specification, rejecting any malformed or non-compliant requests.
    \item \textbf{Threat Detection Patterns:} Implement rules and heuristics to detect suspicious patterns indicative of tool poisoning, command injection, or data exfiltration attempts within tool descriptions, parameters, and responses. Create these rules based on real-world scenarios (example db: \url{https://www.promptfoo.dev/lm-security-db}).
    \item \textbf{Rate-Limiting and Anti-Automation:} Apply granular rate limiting based on source IP, authenticated user, and specific MCP endpoints to mitigate DoS and brute-force attacks. Employ bot detection mechanisms.
    \item \textbf{Comprehensive Request Tracing:} Implement distributed tracing (e.g., using OpenTelemetry) to maintain context across authentication, MCP interactions, and tool invocations for effective auditing and incident analysis.
\end{itemize}
\textit{Rationale:} Provides protocol-specific defense against application-layer attacks.

\subsubsection{Secure Containerization and Orchestration}
Deploy MCP servers in hardened containerized environments only when a code execution environment is not present:
\begin{itemize}
    \item \textbf{Immutable Infrastructure:} Utilize read-only container file systems, mounting writable volumes only where strictly necessary and monitored.
    \item \textbf{Restricted Capabilities:} Drop unnecessary Linux capabilities within containers (e.g., CAP\_NET\_ADMIN, CAP\_SYS\_ADMIN).
    \item \textbf{Resource Quotas:} Enforce strict CPU, memory, network I/O, and storage quotas to prevent resource exhaustion and contain misbehaving components.
    \item \textbf{Seccomp and AppArmor/SELinux:} Apply fine-grained seccomp (secure computing mode) profiles to restrict allowed system calls and use AppArmor or SELinux for mandatory access control (MAC) to further limit potential exploit impact.
    \item \textbf{Regular Scanning:} Integrate vulnerability scanning into the Continuous Integration/Continuous Deployment (CI/CD) pipeline for container images.
\end{itemize}
\textit{Rationale:} Minimizes the attack surface and limits the blast radius of a compromised container.

\subsubsection{Host-Based Security Monitoring}
The goal is to create a comprehensive observability framework capable of detecting potential compromises that may evade higher-level security controls, providing a critical safety net in the complex landscape of AI tool integrations. Deploy Endpoint Detection and Response (EDR) or specialized Host-Based Intrusion Detection Systems (HIDS) on underlying hosts/VMs:
\begin{itemize}
    \item \textbf{MCP-Specific Behavioral Rules:} Develop detection rules tailored to anomalous MCP server process behavior, file access patterns, network connections, or tool invocation sequences.
    \item \textbf{File Integrity Monitoring (FIM):} Continuously monitor critical MCP server binaries, configuration files, and tool definitions for unauthorized modifications.
    \item \textbf{Memory Analysis:} Employ techniques to detect in-memory attacks (e.g., code injection, hooking) that may not leave persistent file system traces.
\end{itemize}
\textit{Rationale:} Provides visibility into potential compromises at the host level, detecting threats that bypass network or application controls.

\subsubsection{Enhanced OAuth 2.0+ Implementation}
Secure MCP server authorization using OAuth 2.0+ principles with enhancements:
\begin{itemize}
    \item \textbf{Strong Client and User Authentication:} Mandate robust client authentication methods (e.g., mTLS, JSON Web Token (JWT) assertion) and require Multi-Factor Authentication (MFA) for user authentication, potentially using adaptive/risk-based authentication.
    \item \textbf{Fine-Grained, Scoped Access Tokens:} Issue short-lived access tokens with the narrowest possible scopes (permissions) required for the specific tool invocation, adhering strictly to the principle of least privilege. Consider Rich Authorization Requests (RAR) for more granular authorization data.
    \item \textbf{Audience Restriction:} Ensure tokens are audience-restricted (aud) to the specific MCP resource server or tool being accessed.
    \item \textbf{Sender-Constrained Tokens:} Implement mechanisms like Demonstration of Proof-of-Possession (DPoP) or mTLS token binding to prevent token theft and replay attacks.
    \item \textbf{Regular Key Rotation:} Automate frequent rotation of keys used for token signing and encryption.
\end{itemize}
\textit{Rationale:} Establishes strong, verifiable identity and authorization as the foundation for access decisions.

\subsubsection{Tool and Prompt Security Management}
Tools in the MCP ecosystem are dynamic, potentially executable entities with complex interaction patterns, rather than static code repositories. This section explores comprehensive strategies for vetting, validating, and monitoring tools throughout their lifecycle, treating each tool as a potential vector for sophisticated attacks such as tool poisoning, data exfiltration, and unauthorized system access.

\textbf{Robust Tool Vetting and Onboarding:} Implement a stringent process for approving and onboarding new tools:
\begin{itemize}
    \item \textit{Security Review Checkpoints:} Mandate security reviews, including static analysis (SAST), dynamic analysis (DAST), dependency scanning (Software Composition Analysis - SCA), and potentially manual code review/pentesting for complex or high-risk tools.
    \item \textit{Clear Documentation:} Require comprehensive documentation detailing the tool's purpose, required permissions, data handling practices, interaction patterns, and security considerations.
    \item \textit{Structured Approval Workflow:} Establish a formal approval process involving security, data governance, privacy, legal, and business stakeholders.
    \item \textit{Periodic Recertification:} Require regular review and recertification of deployed tools to ensure ongoing compliance and security.
\end{itemize}
\textit{Rationale:} Prevents the integration of inherently insecure or malicious tools.

\textbf{Content Security Policy for Tool Descriptions} Treat tool descriptions as potentially executable content and apply controls:
\begin{itemize}
    \item \textit{Structured Validation and Sanitization:} Enforce a strict schema for tool descriptions. Sanitize descriptions rigorously, removing or encoding potentially active content (e.g., script tags, unexpected directives) and validating parameter constraints.
    \item \textit{Malicious Pattern Detection:} Use pattern matching (Regular Expressions - RegEx, YARA rules) and semantic analysis to detect known malicious instruction patterns, command injection attempts, or data exfiltration syntax within descriptions.
    \item \textit{Permission Validation:} Cross-reference requested permissions/scopes against the tool's described functionality, flagging excessive or suspicious requests.
    \item \textit{Integrity Verification:} Implement cryptographic signing and verification for tool descriptions stored in registries to prevent tampering after approval.
\end{itemize}
\textit{Rationale:} Directly mitigates the risk of tool poisoning attacks via manipulated descriptions.

\textbf{Advanced Tool Behavior Monitoring and Poisoning Detection} Go beyond static analysis to detect malicious behavior at runtime:
\begin{itemize}
    \item \textit{Behavioral Baselining:} Establish normal operational baselines for each tool (e.g., typical resource usage, network connections, API calls, data access patterns) and alert on significant deviations.
    \item \textit{Dynamic Analysis/Sandboxing:} Execute tools in isolated sandbox environments during testing or even for initial production requests to observe behavior and detect unexpected actions (e.g., unauthorized network calls, file system access).
    \item \textit{AI/ML-Powered Detection:} Employ machine learning models trained on legitimate and malicious tool behaviors/descriptions to identify novel or obfuscated poisoning techniques.
    \item \textit{Security Telemetry Correlation:} Correlate tool activity logs with broader security telemetry (network logs, endpoint logs, authentication logs) to identify components of sophisticated, multi-stage attacks.
\end{itemize}
\textit{Rationale:} Detects sophisticated poisoning attacks that manifest during execution rather than in static descriptions.

\subsection{MCP Client-Side Mitigations}

\subsubsection{Zero-Trust Security Model Implementation}
The Zero Trust security model represents a paradigm shift from traditional perimeter-based security architectures. Rather than assuming trust based on network location, Zero Trust operates on the fundamental principle of "never trust, always verify." In the context of the Model Context Protocol, this approach becomes particularly critical due to the protocol's inherently dynamic and potentially unpredictable interaction model. The traditional notion of a trusted internal network becomes obsolete when dealing with AI systems that can interact with a vast, constantly changing ecosystem of tools and data sources. Zero Trust provides a comprehensive framework for continuously validating every access attempt, ensuring that each interaction is scrutinized regardless of its origin or apparent legitimacy. Assume no implicit trust; continuously verify every access attempt.

\subsubsection{Just-in-Time (JIT) Access Provisioning}
Just-in-Time (JIT) access provisioning challenges traditional resource access management by eliminating standing privileges. In the MCP ecosystem, where tools and resources can be dynamically invoked and may require temporary, highly specific access, JIT becomes a critical security mechanism. This approach minimizes the attack surface by dramatically reducing the window of potential unauthorized access.
\begin{itemize}
    \item \textbf{Dynamic, Time-Limited Access:} Issue temporary access grants or credentials only for the duration required to complete a specific task or session.
    \item \textbf{Context-Aware Access Decisions:} Base access decisions not only on identity but also on context like device posture, location, time-of-day, and behavioral analytics.
    \item \textbf{Purpose-Driven Authorization:} Align requested permissions with the declared purpose of the tool invocation.
    \item \textbf{Real-Time Revocation:} Implement mechanisms for immediate revocation of access if risk factors change or suspicious activity is detected.
    \item For multi-step or multi-tool access, grant access to each tool only when necessary.
\end{itemize}
\textit{Rationale:} Drastically reduces the window of opportunity for attackers by minimizing persistent access.

% --- BEGIN FIGURE 3 (DOUBLE COLUMN) ---
% --- END FIGURE 3 ---

\subsubsection{Continuous Validation and Monitoring}
Continuous validation departs from traditional authentication models that rely on a single, static authentication event. In the complex and dynamic world of MCP interactions, security cannot be a one-time checkpoint but must be an ongoing, real-time process. Continuously reassess trust throughout interactions by:
\begin{itemize}
    \item \textbf{Per-Request Authorization:} Re-validate authorization for every MCP request, not just at the beginning of a session.
    \item \textbf{Behavioral Anomaly Detection:} Apply User and Entity Behavior Analytics (UEBA) to MCP interactions, identifying deviations from established normal patterns for users and tools.
    \item \textbf{Risk-Based Authentication Step-Up:} Trigger requirements for additional authentication factors if the assessed risk level increases during a session (e.g., sensitive tool invocation, anomalous behavior).
    \item \textbf{Session Monitoring and Reassessment:} Continuously monitor session activity and periodically reassess trust based on changing context or risk signals.
\end{itemize}
\textit{Rationale:} Ensures security posture adapts dynamically to changing conditions and potential threats emerging mid-session.

\subsubsection{Cryptographic Verification of Tool Sources and Integrity}
Ensure tools are authentic and unmodified:
\begin{itemize}
    \item \textbf{Mandatory Code Signing:} Require all integrated tools (especially binaries or executables) to be cryptographically signed by trusted entities (internal Certificate Authority (CA) or reputable third parties). Verify signatures before execution.
    \item \textbf{Secure Tool Registry:} Maintain a secure, preferably immutable registry (e.g., using blockchain or signed Merkle trees), of approved tools, their versions, and corresponding cryptographic hashes or signatures.
    \item \textbf{Supply Chain Security:} Implement measures to secure the tool development and distribution pipeline (e.g., Supply chain levels for Software Artifacts (SLSA) framework) to prevent tampering before integration.
\end{itemize}
\textit{Rationale:} Provides strong assurance of tool authenticity and integrity, preventing the execution of compromised or unauthorized code.

\subsubsection{Input/Output Validation Framework}
Rigorous validation of data flowing through MCP is critical.

\textbf{Strict Schema Validation for MCP Messages} Enforce strict schema validation at MCP ingress points:
\begin{itemize}
    \item \textit{Data Type and Format Enforcement:} Validate data types, formats (e.g., regex for specific patterns), range constraints for numerical values, and enumerated values strictly using allowlists.
    \item \textit{Length and Range Constraints:} Apply realistic length limits to strings and range constraints to numerical values to prevent buffer overflows and resource exhaustion.
    \item \textit{Recursive Validation:} Ensure validation is applied recursively to nested data structures within MCP messages.
    \item \textit{Reject Unknown Fields:} Configure validators to reject messages containing fields not defined in the schema to prevent parameter smuggling or pollution.
\end{itemize}
\textit{Rationale:} Provides a fundamental defense against injection attacks and malformed requests.

\textbf{Context-Aware Input Sanitization and Validation} Go beyond structural validation:
\begin{itemize}
    \item \textit{Contextual Escaping/Encoding:} Apply output encoding appropriate to the context where data will be used by the tool (e.g., HTML escaping, SQL parameterization). Note: This should primarily be handled at the tool level, but MCP can provide contextual guidance.
    \item \textit{Input Normalization:} Normalize inputs (e.g., Unicode canonicalization, case normalization) before validation to prevent evasion techniques.
    \item \textit{Semantic Validation:} Where possible, validate that input values are semantically meaningful within the context of the requested operation.
    \item \textit{Cross-Field Consistency Checks:} Implement rules to check for logical consistency between different fields in a request.
\end{itemize}
\textit{Rationale:} Defends against sophisticated injection and evasion techniques missed by basic schema validation.

\subsection{Additional Security Measures for MCPS}

\subsubsection{Output Filtering and Data Leakage Prevention}
Inspect and filter MCP responses before returning them to the client/AI:
\begin{itemize}
    \item \textbf{Integration with Data Loss Prevention (DLP):} Route MCP responses through enterprise DLP solutions to detect and block sensitive data (Personally Identifiable Information (PII), financial data, secrets) based on predefined policies, often via Internet Content Adaptation Protocol (ICAP) or API integration.
    \item \textbf{Pattern-Based Redaction:} Implement pattern matching (RegEx) to identify and redact common sensitive data types (e.g., credit card numbers, social security numbers, Protected Health Information (PHI)) that might appear inadvertently.
    \item \textbf{Response Size Monitoring:} Alert on unusually large responses, which could indicate data exfiltration attempts.
    \item \textbf{Information Disclosure Prevention:} Filter or modify responses to avoid leaking excessive internal system details, error messages, or stack traces.
\end{itemize}
\textit{Rationale:} Prevents MCP from becoming an unintentional channel for data exfiltration or sensitive information disclosure.

\subsubsection{Operational Security for MCP Environments}
Ongoing security practices are vital.

\textbf{Comprehensive Monitoring and Logging}
\begin{itemize}
    \item \textit{Detailed Event Logging:} Log all significant MCP events: authentication success/failure, authorization decisions, tool registration/updates, tool invocations (including parameters, subject to sanitization), responses (metadata/status), errors, configuration changes.
    \item \textit{Centralized Logging:} Aggregate logs from all MCP components, gateways, hosts, and integrated tools into a central Security Information and Event Management (SIEM) system.
    \item \textit{Correlation and Alerting:} Implement SIEM correlation rules specific to MCP threats (e.g., multiple failed auth attempts followed by success, anomalous tool usage patterns, DLP alerts on MCP traffic). Configure real-time alerts for high-priority events.
    \item \textit{Immutable Audit Trails:} Ensure logs are tamper-evident and stored securely with appropriate retention policies meeting compliance and forensic requirements.
\end{itemize}
\textit{Rationale:} Provides visibility for threat detection, incident response, and compliance auditing.

\textbf{Tailored Incident Response Procedures} Develop specific playbooks for MCP-related incidents:
\begin{itemize}
    \item \textit{MCP Incident Categories:} Define classifications for incidents like suspected tool poisoning, unauthorized access via MCP, data exfiltration, MCP server compromise, DoS attacks.
    \item \textit{Response Playbooks:} Create step-by-step procedures for each category, covering: Containment (Rapidly isolating affected MCP components or quarantining suspicious tools), Eradication (Removing malicious code or configurations), Recovery (Securely restoring MCP services), Evidence Preservation (Collecting relevant logs and system state for forensic analysis), Communication (Notifying relevant stakeholders).
    \item \textit{Post-Incident Analysis:} Conduct thorough root cause analysis (RCA) and update security controls accordingly.
\end{itemize}
\textit{Rationale:} Enables rapid and effective response to minimize the impact of MCP security incidents.

\textbf{Threat Intelligence Integration} Stay informed about evolving threats:
\begin{itemize}
    \item \textit{MCP/AI-Specific Feeds:} Subscribe to threat intelligence feeds focusing on AI security, API threats, and vulnerabilities in MCP implementations or common dependencies. Track relevant Tactics, Techniques, and Procedures (TTPs) and Indicators of Compromise (IoCs).
    \item \textit{Indicator Sharing:} Participate in relevant Information Sharing and Analysis Centers (ISACs) or communities focused on AI/API security.
    \item \textit{Proactive Threat Hunting:} Use threat intelligence (TTPs, IoCs) to proactively hunt for signs of compromise within the MCP environment.
    \item \textit{Intelligence-Driven Testing:} Incorporate insights from threat intelligence into penetration testing and red teaming exercises targeting the MCP deployment.
\end{itemize}
\textit{Rationale:} Ensures defenses remain effective against the latest attacker techniques targeting MCP and related systems.

\textbf{Automated Security Operations (SecOps)} Leverage automation for efficiency and consistency:
\begin{itemize}
    \item \textit{Automated Patching:} Implement automated workflows for testing and deploying security patches for MCP components and underlying infrastructure.
    \item \textit{Configuration Drift Detection and Remediation:} Use configuration management tools (e.g., Ansible, Terraform) and security posture management tools to detect and automatically correct deviations from secure baselines.
    \item \textit{Automated Containment:} Configure Security Orchestration, Automation, and Response (SOAR) platforms to automatically trigger containment actions (e.g., block IP, disable tool, isolate container) based on high-confidence alerts from the SIEM.
    \item \textit{Automated Access Reviews:} Implement automated workflows for periodic user and tool access recertification.
\end{itemize}
\textit{Rationale:} Improves response times, reduces human error, and ensures consistent application of security policies.

\subsubsection{Security Requirements for Hosting a Public Anthropic MCP Server}
In order to host a public secure MCP Server, there's a need to establish a robust security framework that involves key security practices aimed at reducing the server's attack surface and restricting the capabilities of a malicious actor.
\begin{itemize}
    \item \textbf{Agent Boundaries and State Isolation:} The sessions of individual agents must be isolated and kept separate from other sessions, preventing cross-session or cross-agent interference.
    \item \textbf{Input Validation and Sanitization:} Validate and sanitize inputs (e.g., user prompts, tool parameters) passed to the server in order to prevent prompt injection and parameter pollution attempts.
    \item \textbf{Operating System Hardening:} Deploy a minimal operating system comprising solely needed components and services. Strengthen the kernel by implementing appropriate security parameters, deactivating unnecessary kernel modules, and employing security-enhancing tools such as SELinux or AppArmor for required access control.
    \item \textbf{Service and Port Management:} Identify and deactivate all services and network ports that are not essential for the functioning of the MCP server. Conduct regular audits of active services to verify that only permitted services are operational, thus reducing vulnerability to threats.
    \item \textbf{Secure Remote Access:} Prohibit direct root login using SSH and require SSH key-based authentication. Alter the default SSH port to a non-standard designation and enforce IP allowlisting to limit SSH access to authorized networks. To enhance security, utilize a VPN for remote access from untrusted networks.
    \item \textbf{Strict Firewall Configurations:} Configure firewalls according to a "deny all by default" approach, periodically assess and refine firewall rules, and implement Web Application Firewalls (WAFs) for enhanced security.
    \item \textbf{Employ Intrusion Detection and Prevention Systems (IDPS):} Utilize IDPS to oversee network traffic and issue alerts for anomalous behavior.
    \item \textbf{Enforce MFA for High-Privilege Access:} Mandate MFA for administrator logins and access to sensitive systems, employing mechanisms such as authenticator applications or hardware tokens.
\end{itemize}

\subsubsection{Security Requirements for Multi-MCP Server Deployments}
As organizations deploy multiple MCP servers to enhance AI capabilities across their ecosystem, specific security considerations become essential. Multi-server deployments introduce unique challenges related to isolation, authorization, and trust boundaries that must be addressed through a comprehensive security framework.

\textbf{Hosting Requirements}
\begin{itemize}
    \item \textit{Containerized Deployment:} Implement Docker-based containerization for MCP servers to ensure consistent security configurations across deployments and prevent environment conflicts as highlighted by Docker's collaboration with Anthropic, which addresses challenges like dependency conflicts, cross-platform consistency, and complex setup requirements.
    \item \textit{Secure Remote Hosting:} When using cloud-based or remote MCP server hosting, implement dedicated security controls that isolate server instances and enforce strict access boundaries between environments similar to Cloudflare's approach for remote MCP server hosting with built-in OAuth support and enhanced accessibility \cite{ssojet2025cloudflare}.
    \item \textit{Secrets Management:} Implement a centralized secrets management system for managing authentication credentials across multiple MCP servers using a solution that injects API keys or other secrets on-the-fly rather than storing them in files that could be leaked.
    \item \textit{Resource Isolation:} Configure resource quotas and limitations for each MCP server to prevent resource exhaustion attacks from compromising the availability of other servers in the deployment.
\end{itemize}

\textbf{MCP Server Isolation}
\begin{itemize}
    \item \textit{Network-Level Segmentation:} Implement strict network policies that limit communication between MCP servers and enforce explicit allow rules for necessary interactions to prevent lateral movement in the event one server is compromised, as MCP servers represent high-value targets storing authentication tokens for multiple services.
    \item \textit{Session State Isolation:} Ensure complete isolation of agent sessions between different MCP servers to prevent cross-session interference or data leakage, preventing scenarios where one compromised server could potentially access data from another server's sessions.
    \item \textit{Environmental Separation:} Deploy distinct MCP servers for different security contexts (e.g., development, testing, production) with no shared authentication or access mechanisms between environments.
\end{itemize}

\textbf{Trust and Safety}
\begin{itemize}
    \item \textit{Capability Discovery Controls:} Implement controls around capability discovery to ensure MCP clients can only discover and access authorized tools from approved servers similar to Cloudflare's MCPClientManager implementation that handles capability discovery with proper namespacing when connecting to multiple MCP servers \cite{ssojet2025cloudflare}.
    \item \textit{Server Authentication:} Implement cryptographic verification of MCP server identities to prevent server spoofing attacks where malicious servers could pose as legitimate ones, mitigating risks where attackers could create their own MCP server instances using stolen credentials.
    \item \textit{Tool Validation Framework:} Establish a centralized validation framework for approving and auditing tools across all MCP servers, with stricter validation requirements for tools with elevated privileges or access to sensitive data.
    \item \textit{Cross-Server Monitoring:} Implement coordinated security monitoring across all MCP server instances to detect patterns of suspicious activity that may span multiple servers or clients.
\end{itemize}

% IEEE uses tables* for tables spanning both columns
\begin{table*}[!t]
\renewcommand{\arraystretch}{1.3} % Increase row spacing
\caption{MCP Security Threats and Mitigation Controls}
\label{tab:threats_mitigations}
\centering
% Use tabularx for better column width management
\begin{tabularx}{\textwidth}{@{}l X X@{}} % X column type allows text wrapping
\toprule
\textbf{Threat Category} & \textbf{Description} & \textbf{Key Controls} \\
\midrule
Tool Poisoning & Malicious manipulation of tool descriptions or parameters to induce unintended or harmful AI model actions & \begin{tabular}[t]{@{}l@{}}• Content Security Policy for tool descriptions\\ • Tool behavior monitoring\\ • Semantic analysis of tool descriptions\\ • Sandboxed execution\end{tabular} \\
\addlinespace % Add some space between rows
Data Exfiltration & Unauthorized extraction of sensitive data through compromised tools or manipulated MCP responses & \begin{tabular}[t]{@{}l@{}}• Output filtering with DLP integration\\ • Response size monitoring\\ • Pattern-based redaction\\ • Anomaly detection\end{tabular} \\
\addlinespace
Command and Control (C2) / Update Mechanism Compromise & Establishment of covert channels via compromised MCP servers or tools / Insertion of persistent backdoors through compromised MCP server or tool update channels & \begin{tabular}[t]{@{}l@{}}• Network segmentation / Egress filtering\\ • Behavioral analysis\\ • Tool isolation\\ • Cryptographic verification / Secure tool registry\\ • Supply chain security\\ • File integrity monitoring\end{tabular} \\
\addlinespace
Identity/Access Control Subversion & Exploitation of authentication or authorization flaws to gain unauthorized access & \begin{tabular}[t]{@{}l@{}}• Enhanced OAuth implementation\\ • JIT access provisioning\\ • MFA for privileged access\\ • Continuous validation\end{tabular} \\
\addlinespace
Denial of Service (DoS) & Overloading MCP servers or dependent resources through excessive requests & \begin{tabular}[t]{@{}l@{}}• Rate limiting\\ • Resource quotas\\ • Anti-automation\\ • Redundancy\end{tabular} \\
\addlinespace
Insecure Configuration & Exploitation of misconfigurations in MCP servers or network settings & \begin{tabular}[t]{@{}l@{}}• Configuration hardening / Secure defaults\\ • Automated drift detection\\ • Regular audits\end{tabular} \\
\bottomrule
\end{tabularx}
\end{table*}

\section{Implementation Strategies for Enterprise Environments}

Choosing the right deployment pattern depends on existing infrastructure, risk tolerance, and operational capabilities.

\subsection{Secure MCP Deployment Patterns}

\begin{itemize}
    \item \textbf{Pattern 1: Dedicated Security Zone Architecture}
        \begin{itemize}
            \item \textit{Description:} Isolate all MCP components (servers, databases, supporting services) within a dedicated, highly restricted network segment with strict firewall rules, dedicated monitoring, and potentially separate Identity and Access Management (IAM).
            \item \textit{Pros:} Strong isolation, clear security boundaries, easier compliance demonstration for high-security environments.
            \item \textit{Cons:} Can increase complexity and operational overhead, potentially creates infrastructure silos.
            \item \textit{Suitable for:} Organizations with stringent security/compliance needs (finance, healthcare), mature network segmentation practices.
        \end{itemize}
    \item \textbf{Pattern 2: API Gateway-Centric Integration}
        \begin{itemize}
            \item \textit{Description:} Place MCP servers behind an existing enterprise API gateway, leveraging the gateway for authentication, authorization, rate limiting, WAF capabilities, and unified logging/monitoring.
            \item \textit{Pros:} Leverages existing investments, consistent policy enforcement across all APIs, potentially faster deployment.
            \item \textit{Cons:} Security depends heavily on gateway capabilities and configuration; MCP-specific logic might still be needed beyond the gateway.
            \item \textit{Suitable for:} Organizations with mature API management platforms and a desire for centralized API governance.
        \end{itemize}
    \item \textbf{Pattern 3: Containerized Microservices within Orchestration}
        \begin{itemize}
            \item \textit{Description:} Deploy MCP components as microservices within a container orchestration platform (e.g., Kubernetes). Leverage platform features like network policies, secrets management, service meshes, and automated scaling/healing.
            \item \textit{Pros:} Operational flexibility, scalability, resilience, fine-grained control via platform features (e.g., Kubernetes NetworkPolicies, Istio).
            \item \textit{Cons:} Requires container orchestration expertise, security relies on correct platform configuration.
            \item \textit{Suitable for:} Organizations utilizing cloud-native architectures and container orchestration.
        \end{itemize}
\end{itemize}

\subsection{Integration with Enterprise Security Ecosystem}

MCP security cannot exist in a vacuum. Integration is key.
\begin{itemize}
    \item \textbf{Identity and Access Management (IAM):} Integrate with enterprise IAM (e.g., Azure AD, Okta) for user authentication (Single Sign-On - SSO), centralized identity governance, and potentially leveraging group memberships or attributes for coarse-grained authorization. OAuth/OpenID Connect (OIDC) federation is crucial here.
    \item \textbf{Security Information and Event Management (SIEM):} Forward all MCP logs (Section 3.3.2) to the enterprise SIEM (e.g., Splunk, QRadar, Sentinel) for correlation with other security data, centralized alerting, and unified incident investigation.
    \item \textbf{Data Loss Prevention (DLP):} Integrate MCP output filtering (Section 3.3.1) with enterprise DLP solutions via ICAP or API integrations to enforce consistent data protection policies across all egress channels \cite{guilherme2025llm}.
    \item \textbf{Secrets Management:} Utilize enterprise secrets management solutions (e.g., HashiCorp Vault, AWS Secrets Manager) for securely storing and managing API keys, certificates, and credentials used by MCP servers and tools.
\end{itemize}
\textit{Rationale:} Ensures MCP security aligns with overall enterprise security posture, policies, and tooling, providing holistic visibility and control.

\section{Limitations and Implementation Challenges}

While the proposed framework provides a comprehensive approach, organizations should be aware of inherent limitations and potential implementation challenges:
\begin{itemize}
    \item \textbf{Complexity:} Implementing the full suite of controls requires significant security expertise, potentially new tooling investments, and ongoing operational resources.
    \item \textbf{Performance Overhead:} Certain security measures, such as deep packet inspection, complex cryptographic operations (e.g., DPoP), and intensive real-time monitoring, can introduce latency or performance overhead that needs careful management, especially in low-latency applications.
    \item \textbf{Tool Ecosystem Maturity and Vetting:} The effectiveness of the framework relies partially on the security posture of third-party tools integrated via MCP. Vetting external tools thoroughly can be challenging and resource-intensive, and the security quality of available tools may vary significantly.
    \item \textbf{Dynamic Threat Landscape:} AI models, tools, and attack techniques evolve rapidly. Security controls must be continuously reviewed, updated, and adapted to remain effective against emerging threats.
    \item \textbf{Empirical Validation Gap:} As a relatively new protocol, there is limited large-scale, publicly available data on real-world MCP attacks and the measured effectiveness of specific countermeasures. The validation in this paper is primarily conceptual and scenario-based.
    \item \textbf{Usability and Developer Experience:} Overly stringent security controls, if not implemented carefully, could potentially hinder developer productivity or negatively impact the usability of AI applications relying on MCP. Finding the right balance is crucial.
    \item \textbf{Policy Enforcement Consistency:} Ensuring consistent application of fine-grained security policies across a potentially large and diverse set of tools and MCP instances can be operationally complex.
\end{itemize}
Acknowledging these challenges is crucial for realistic planning and successful implementation of MCP security.

\section{Future Research Directions}

MCP security is an evolving field. Key areas for future research include:
\begin{itemize}
    \item \textbf{AI-Driven Security for MCP:} Researching the use of AI/ML specifically for defending MCP, such as: Advanced, context-aware tool poisoning detection models capable of understanding semantic manipulation; Reinforcement learning for adaptive MCP security policy tuning based on observed threats; AI-powered generation and validation of secure MCP configurations and tool manifests.
    \item \textbf{Confidential Computing for MCP:} Investigating the application of confidential computing techniques (e.g., secure enclaves like Intel SGX, AMD SEV) to protect MCP server processes and sensitive context data even from compromised host operating systems or cloud providers.
    \item \textbf{Standardization of MCP Security Extensions:} Developing standardized extensions to the MCP protocol itself to incorporate security features like enhanced metadata for tool vetting (e.g., security attestations), standardized security event formats for improved SIEM integration, or built-in integrity mechanisms for protocol messages.
    \item \textbf{Cross-Protocol Security in AI Ecosystems:} Analyzing security implications arising from interactions between MCP and other AI/ML protocols or frameworks (e.g., ML Ops pipelines, federated learning protocols, other agent communication standards).
    \item \textbf{Measurable Security Metrics for MCP:} Developing standardized metrics and methodologies for quantitatively assessing the security posture of MCP deployments and the effectiveness of specific controls.
\end{itemize}

\section{Conclusion}

The Model Context Protocol offers powerful capabilities for extending AI systems but introduces significant security challenges that require proactive and sophisticated mitigation. Simply adopting standard API security practices is insufficient. This paper has presented a comprehensive, multi-layered security framework specifically tailored for MCP, emphasizing defense-in-depth, Zero Trust principles, rigorous tool vetting, continuous monitoring, and robust input/output validation.

We provided detailed implementation strategies, operational guidelines, and reference patterns designed to be actionable for security practitioners building or managing MCP deployments in enterprise environments. The framework aids prioritization through its structure and the inclusion of qualitative risk assessments (Table \ref{tab:threats_mitigations}). While the threat landscape will continue to evolve, and implementation presents challenges (Section V), implementing the framework described herein---integrating network, application, host, data, and identity controls---provides a strong foundation for securely leveraging MCP.

Organizations must treat MCP security as a critical priority from the outset of any implementation, integrating it into their overall AI governance strategy. By adopting a security-first mindset, implementing robust technical controls, establishing strong operational practices, acknowledging limitations, and staying abreast of emerging threats and research, enterprises can confidently harness the transformative potential of MCP while effectively managing the associated risks.

\bibliographystyle{IEEEtran}

\bibliography{IEEEabrv,references}

\end{document}